%% file: smmpaper.tex
\documentclass[
aps,
prd,
longbibliography,
reprint,
floatfix,
preprintnumbers,
superscriptaddress
]{revtex4-2}

\usepackage[T1]{fontenc}

\usepackage{iftex}
\usepackage{hyperref}
\ifpdf
  \hypersetup{colorlinks=true}
\fi
\usepackage{xspace}
\usepackage{graphicx}
\usepackage{lineno}
\newcommand{\fig}[1]{Fig.~\ref{fig:#1}}
\newcommand{\tab}[1]{Table~\ref{tab:#1}}
\newcommand{\sect}[1]{Section~\ref{sec:#1}}

\newcommand{\regularfigurescale}{0.975\columnwidth}

\newcommand{\citeuncat}{S190510g,S190718y,S190901ap,S190910d,S190910h,%
S190923y,S190930t,S191105e,S191109d,S191129u,S191204r,S191205ah,S191213g,%
S191215w,S191216ap,S191222n,S200105ae,S200112r,S200114f,S200115j,S200128d,%
S200129m,S200208q,S200213t,S200219ac,S200224ca,S200225q,S200302c,S200311bg,%
S200316bj}

\newcommand{\figureone}
{
\begin{figure}
\centering

\includegraphics[width=\regularfigurescale]{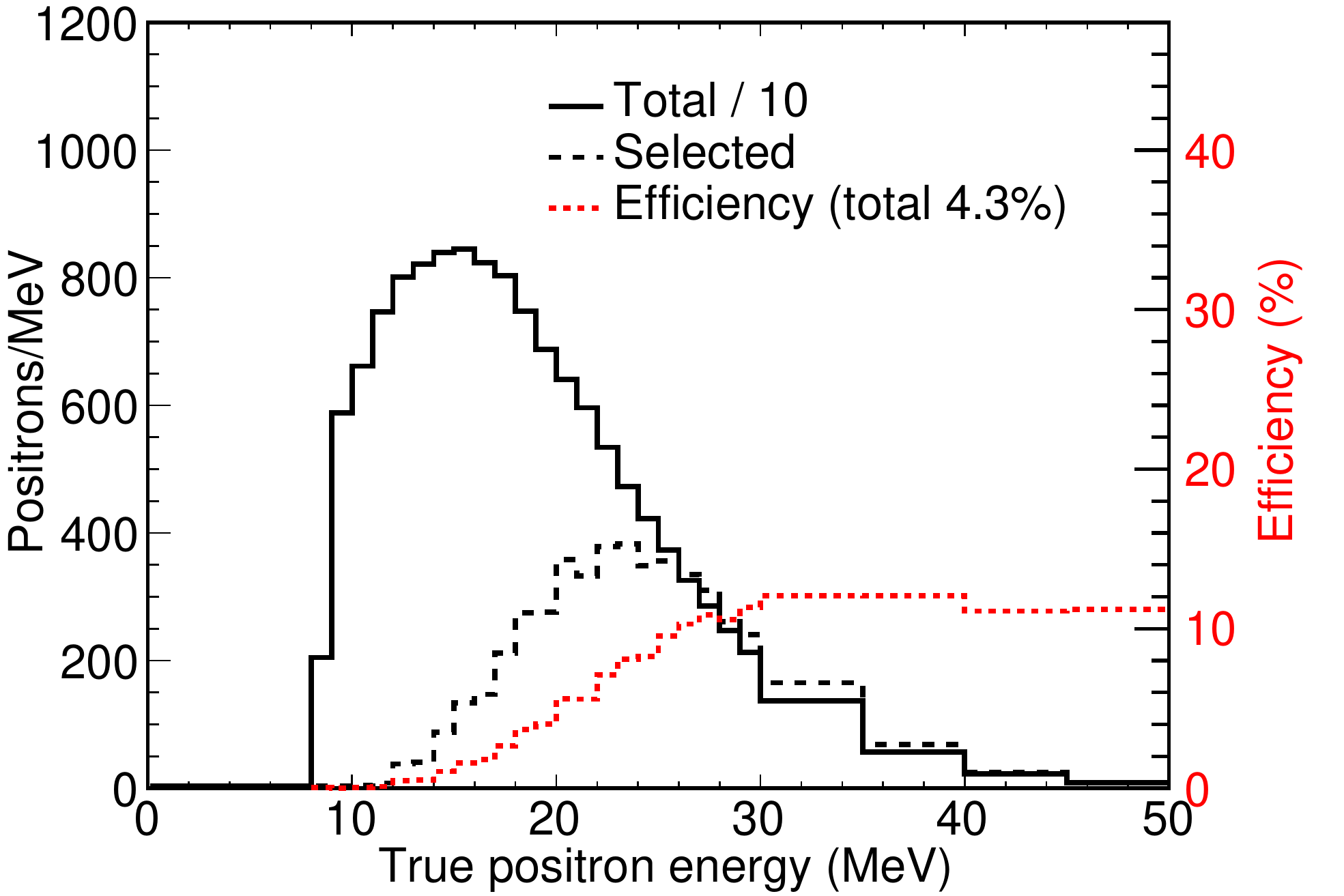}

\includegraphics[width=\regularfigurescale]{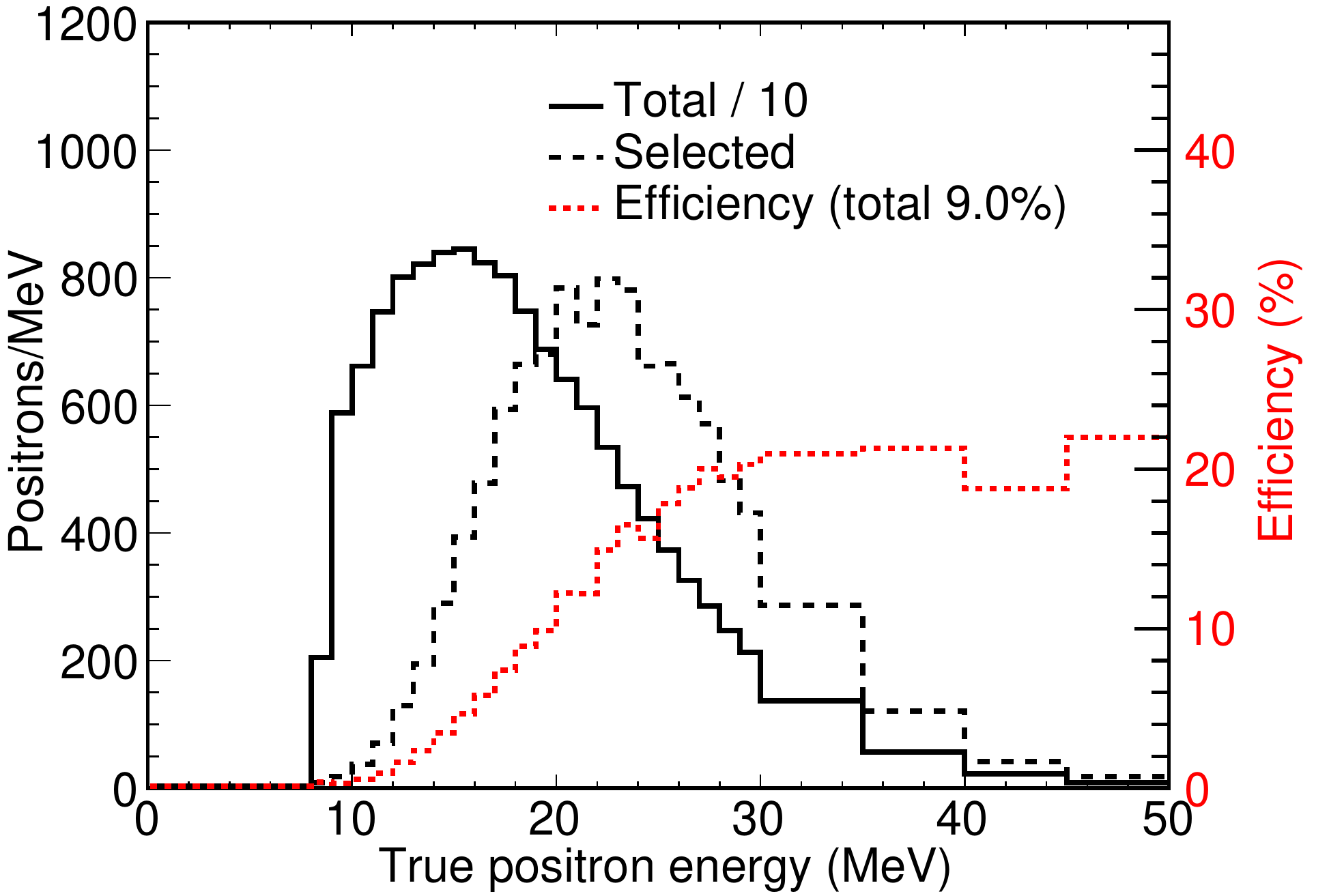}

\includegraphics[width=\regularfigurescale]{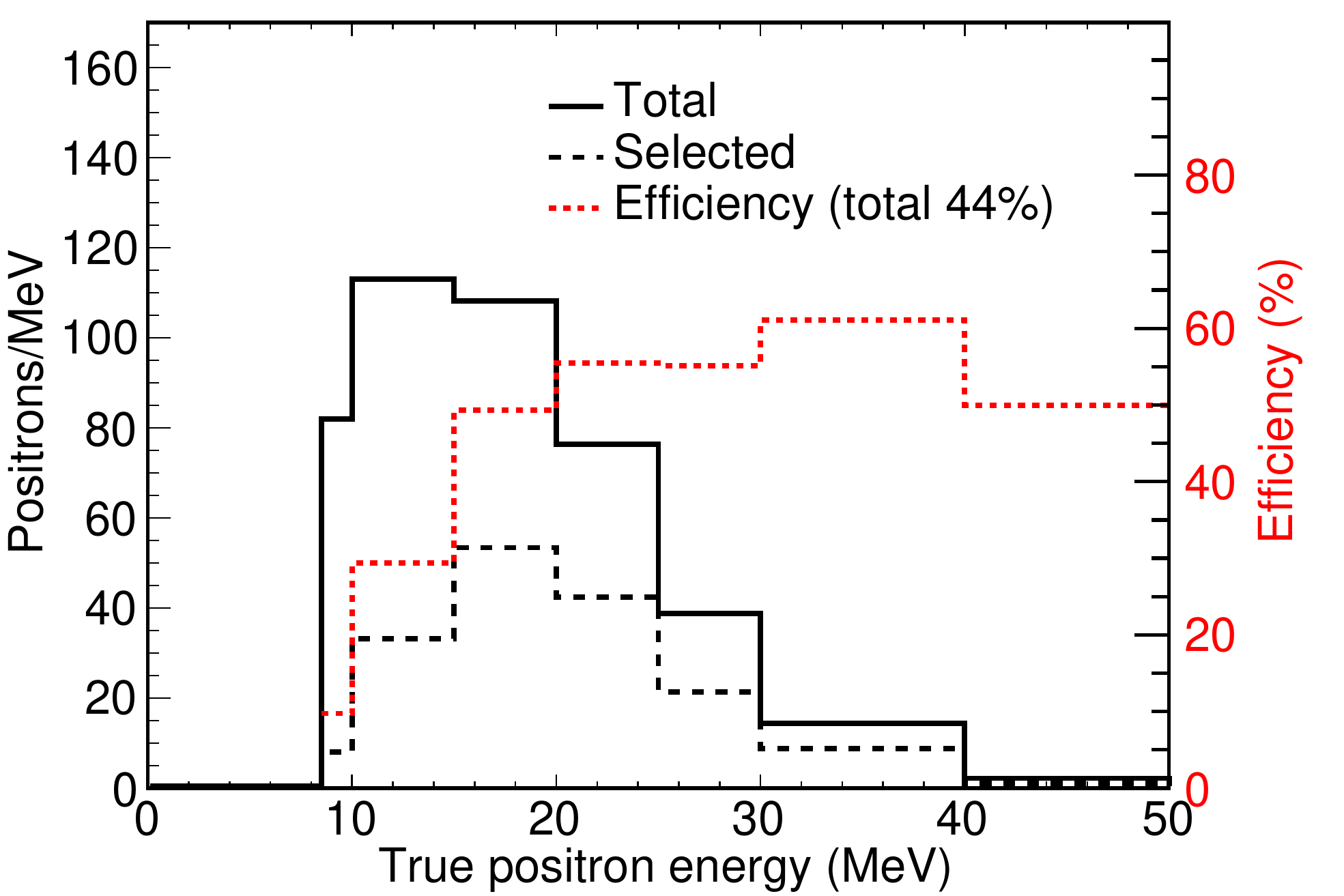}

\caption{Top (middle, bottom): FD continuous readout (FD pulser, ND) total and
selected positron spectra (left axis) and positron selection
efficiency (right axis) as a function of energy. Each plot shows a simulated 9.6\ms supernova
at 1\,kpc.  The total efficiencies, integrated over neutrino energy, are shown in the legends and assume a 9.6\ms supernova; total
efficiencies are higher for the 27\ms model.}

\label{fig:positroneff}
\end{figure}
}

\newcommand{\figuretwo}
{
\begin{figure}
\centering

\includegraphics[width=\regularfigurescale]{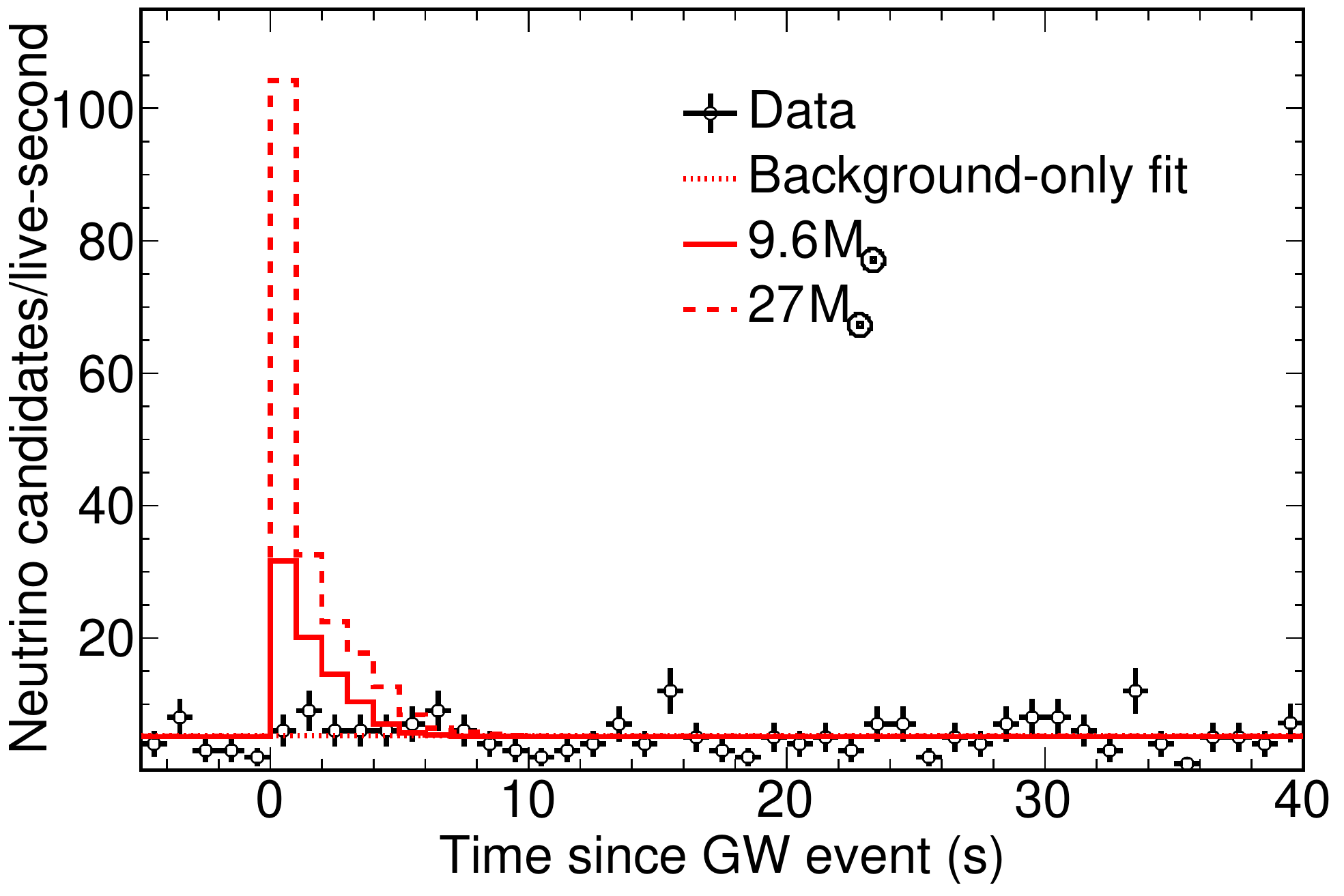}

\includegraphics[width=\regularfigurescale]{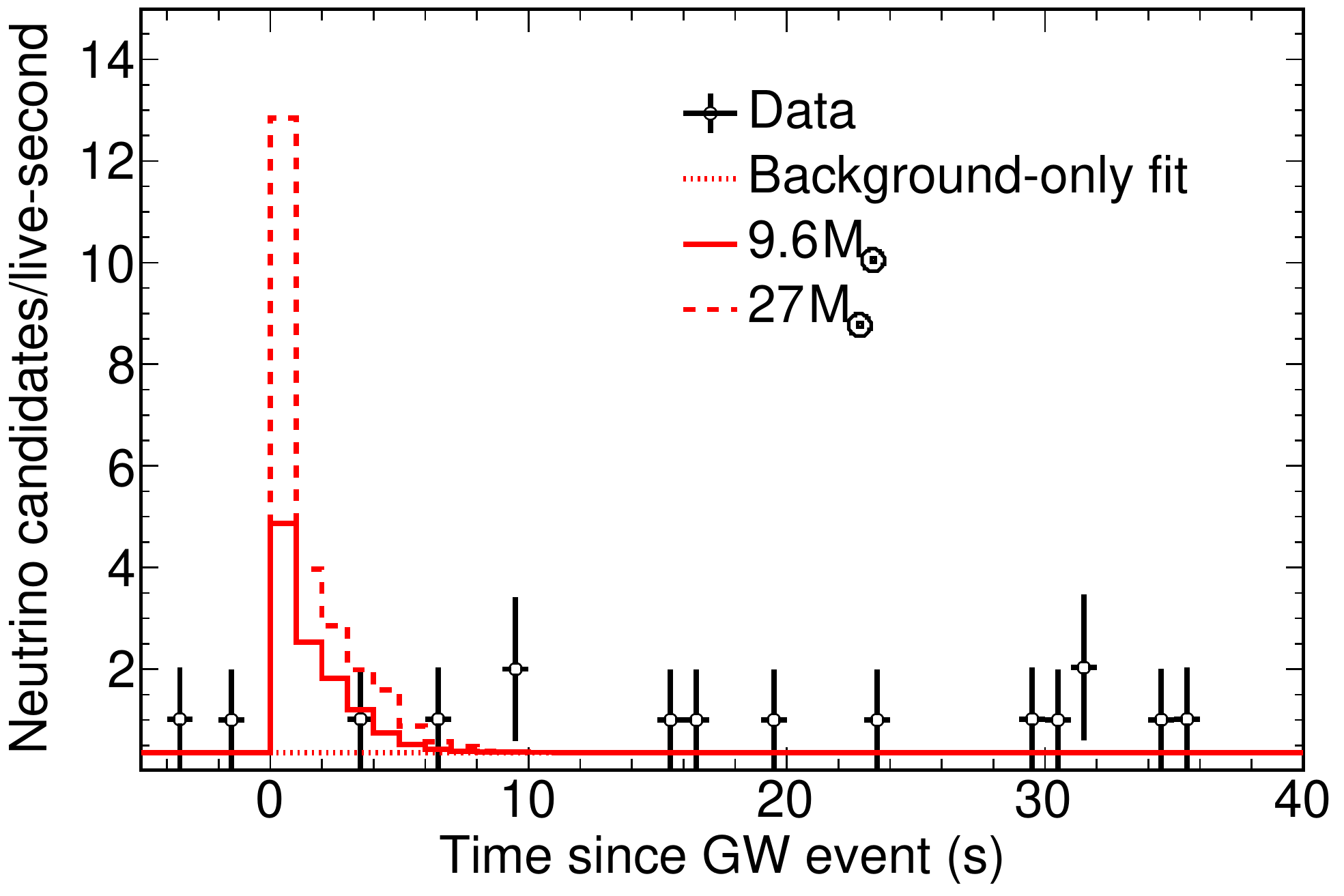}

\caption{A typical GW event with both FD (top) and ND (bottom) continuous readout, S200213t.
The two supernova models are shown, normalized
to 10\,kpc.  The number of neutrino candidates per second is corrected for livetime, which
is slightly under 100\% in the ND because of beam removal, and in the final
bins because readout ends at 39.86\,s.
The limits set are weaker
than the median case because of a slight excess in the 0--5\,s bins in the statistically-dominant FD.}

\label{fig:timedists}
\end{figure}
}

\newcommand{\tabheader}{
\begin{tabular}{l c c c c c c}
\hline
\hline
& & & \multicolumn{2}{c}{Fluence} & \multicolumn{2}{c}{Distance} \\
Name & ND & FD &
$\mathrm{SN_{27\odot}}$ & $\mathrm{SN_{9.6\odot}}$ & 
$\mathrm{SN_{27\odot}}$ & $\mathrm{SN_{9.6\odot}}$ \\
\hline
}

\newcommand{\untrig}{Untriggered}
\newcommand{\ten}{Untriggered}
\newcommand{\full}{45.0\,s}

\newcommand{\zz}{\phantom{00}}
\newcommand{\z}{\phantom{0}}
\newcommand{\pz}{\phantom{.0}}

\newcommand{\tableone}
{
\begin{table*}

\caption{Summary of NOvA data taking during GW
events~\cite{ligocat,ligo2020cat,\citeuncat} and 90\% C.L. limits.  The fluence limits on
the two supernova models are 
in units of $10^{10}\mathrm{\,cm^{-2}}$.  The distance limits are in kiloparsecs. When continuous data was read
out in response to an LVC trigger, the number of seconds read is given for each
detector. Otherwise (``untriggered''), pulser data is used in the case of the FD, and the ND is not used. 
In some cases one or both detectors were not running (``no data'') and in two cases
the FD was running, but not taking good data (``bad''). Events above the line have been considered by NOvA
before; above and below the line events are arranged chronologically. }

\label{tab:events}

\begin{center}
\tabheader

GW150914 & \untrig & Bad         & ---\phantom{0}    & \phantom{0}---   &  \phantom{0}--- & --- \\
GW151012 & \untrig & No data     & ---\phantom{0}    & \phantom{0}---   &  \phantom{0}--- & --- \\
GW151226&\untrig&\ten&110\pz&190 & \zz 9&\z 6\pz\\
GW170104&\untrig&\ten&300\pz&500 & \zz 6 & \z 3.4 \\
GW170608&\untrig&\ten&400\pz&700 & \zz 5&\z 2.9\\
GW170729&\untrig&\ten&240\pz&400&\zz 6&\z 4\pz \\
GW170809&\untrig&\ten&110\pz&190 &\zz 9&\z 6\pz\\
GW170814&\untrig&\ten&120\pz&200 &\zz 9&\z 5\pz\\
GW170817&\untrig&\ten&110\pz&190 &\zz 9&\z 6\pz\\
GW170818&\untrig&\ten&180\pz&330 &\zz 7&\z 4\pz\\
GW170823&\untrig&\ten&260\pz&500 &\zz 6&\z 3.5\\
GW190408\_181802 & No data & No data & ---\phantom{0} & \phantom{0}--- &  \phantom{0}--- & --- \\
GW190412&\untrig&\ten&170\pz&280 & \zz 7&\z 4\pz\\
GW190421\_213856&\untrig&\ten&210\pz&400 &\zz 7&\z 4\pz\\
GW190425&\untrig&\ten&120\pz&190 &\zz 9&\z 5\pz\\
GW190426\_152155&44.7\,s&\ten&\z 13\pz&\z 19 &\z 27& 17\pz\\
GW190503\_185404&\untrig&\ten&150\pz&270 &\zz 8&\z 5\pz\\
S190510g&\untrig&\ten&170\pz&280 &\zz 7&\z 4\pz\\
GW190512\_180714&\untrig&\ten&190\pz&330 &\zz 7&\z 4\pz\\
GW190513\_205428&24.7\,s&\ten&\z 14\pz&\z 20 &\z 26& 17\pz\\
GW190517\_055101&\untrig&\ten&120\pz&200 &\zz 9&\z 5\pz\\
GW190519\_153544&\untrig&\ten&140\pz&250 &\zz 8&\z 5\pz\\
GW190521&\full&\full&\zz 6\pz&\z 10 &\z 40& 24\pz\\
GW190521\_074359&\untrig&\ten&170\pz&280 &\zz 7&\z 4\pz\\
GW190602\_175927&\full&\full&\zz 6\pz&\z 12 &\z 40& 22\pz\\
GW190630\_185205&\full&\full&\zz 5\pz&\zz 9 &\z 40& 25\pz\\
GW190701\_203306&\full&\full&\zz 6\pz&\z 11 &\z 40& 23\pz\\
GW190706\_222641&\full&17.5\,s&\zz 2.5&\zz 5 &\z 60& 35\pz\\
GW190707\_093326&\untrig&\ten&220\pz&400 &\zz 6&\z 4\pz\\
\hline
GW190413\_052954&\untrig&\ten&170\pz&280&\zz 7&\z 4\pz\\
GW190413\_134308&\untrig&\ten&160\pz&270&\zz 8&\z 5\pz\\
GW190424\_180646&\untrig&\ten&140\pz&240&\zz 8&\z 5\pz\\
GW190514\_065416&\untrig&\ten&280\pz&500&\zz 6&\z 3.5\\
GW190527\_092055&\untrig&\ten&140\pz&240&\zz 8&\z 5\pz\\
GW190620\_030421&\untrig&\ten&270\pz&400&\zz 6&\z 4\pz\\
GW190708\_232457&\untrig&\ten&150\pz&270&\zz 8&\z 5\pz\\
S190718y&18.3\,s&\ten    &\z 17\pz&\z 23&\z 23& 16\pz\\
GW190719\_215514&\untrig & Bad & ---\phantom{0}& \phantom{0}--- &  \phantom{0}--- & --- \\
GW190720\_000836&\full&\full&\zz 4\pz&\zz 6&\z 50& 31\pz\\
GW190727\_060333&\full&\full&\zz 5\pz&\zz 9&\z 40& 25\pz\\

\hline
\hline
\end{tabular}
\end{center}
\end{table*}
}

\newcommand{\tableonetwo}
{
\begin{table*}

\caption{Continuation of \tab{events}.}

\label{tab:events2}

\begin{center}

\tabheader

GW190728\_064510&\full&29.6\,s&\zz 3.2&\zz 5&\z 50& 33\pz\\
GW190731\_140936&\untrig&\ten&210\pz&400&\zz 7&\z 4\pz\\
GW190803\_022701&\untrig&\ten&140\pz&230&\zz 8&\z 5\pz\\
GW190814&\full&\ten&\z 14\pz&\z 22&\z 25& 16\pz\\
GW190828\_063405&\full&18.1\,s&\zz 6\pz&\z 10&\z 40& 23\pz\\
GW190828\_065509&\full&\ten&\z 16\pz&\z 21&\z 24& 16\pz\\
S190901ap&\full&\full&\zz 3.1&\zz 6&\z 60& 30\pz\\
GW190909\_114149&\untrig&\untrig& 110\pz & 190 & \zz 9 &\z 5\pz \\
S190910d&\full&\full&\zz 4\pz&\zz 7&\z 50& 29\pz\\
S190910h&\full&\full&\zz 2.7&\zz 5&\z 60& 32\pz\\
GW190910\_112807&\untrig&\ten&120\pz&190&\zz 9&\z 6\pz\\
GW190915\_235702&\full&\full&\zz 3.0&\zz 6&\z 60& 31\pz\\
S190923y&\full&\full&\zz 3.2&\zz 6&\z 50& 32\pz\\
GW190924\_021846&\full&\full&\zz 4\pz&\zz 7&\z 50& 28\pz\\
GW190929\_012149&\untrig&\ten&200\pz&340&\zz 7&\z 4\pz\\
GW190930\_133541&\full&\full&\zz 7\pz&\z 13&\z 40& 21\pz\\
S190930t&\full&\full&\zz 5\pz&\z 10   &\z 40& 23\pz\\
S191105e&\untrig &\ten&180\pz&310     &\zz7&\z 4\pz\\
S191109d&\full&\full&\zz 5\pz&\zz 8   &\z 40& 26\pz\\
S191129u&\untrig &\ten&230\pz&400     &\zz6&\z 4\pz\\
S191204r&\untrig &\ten&300\pz&500     &\zz6&\z 3.4\\
S191205ah&\full&\full&\zz 2.7&\zz 6   &\z 60& 32\pz\\
S191213g&\full&\full &\zz 3.4&\zz 7   &\z 50& 29\pz\\
S191215w&\full&\full&\zz 4\pz&\zz 7   &\z 50& 29\pz\\
S191216ap&\full&29.5\,s&\zz2.7&\zz5   &\z60& 35\pz\\
S191222n&\full&\full&\zz 4\pz&\zz 7   &\z 50& 29\pz\\
S200105ae&\untrig&\ten&230\pz&400     &\zz6&\z 4\pz\\
S200112r&\full&No data &\z 16\pz&\z 23&\z 24& 16\pz\\
S200114f&\full&\full&\zz 9\pz&\z 15   &\z 32& 19\pz\\
S200115j&\full&\full &\zz 2.1&\zz 4   &\z 70& 40\pz\\
S200128d&\full&\full&\zz 5\pz&\zz 8   &\z 50& 26\pz\\
S200129m&\full&\full &\zz 3.2&\zz 6   &\z 50& 32\pz\\
S200208q&\full&\full&\zz 5\pz&\zz 7   &\z 40& 28\pz\\
S200213t&\full&\full&\zz 5\pz&\z 10   &\z 40& 23\pz\\
S200219ac&\untrig&\ten&190\pz&300     &\zz7&\z 4\pz\\
S200224ca&\full&No data&\z 22\pz&\z 29&\z 21& 14\pz\\
S200225q&\full&\full &\zz 3.4&\zz 6   &\z 50& 30\pz\\
S200302c&\full&\full&\zz 4\pz&\zz 8   &\z 50& 26\pz\\
S200311bg&\full&No data&\z 16\pz&\z 21&\z 24& 16\pz\\
S200316bj&\full&\full&\zz 2.9&\zz 5   &\z 60& 33\pz\\

\hline
\hline
\end{tabular}
\end{center}
\end{table*}
}

\newcommand{\flusym}{F}
\newcommand{\flu}[1]{$\flusym < #1\times 10^{10}\,\mathrm{cm}^{-2}$\xspace}
\newcommand{\flubare}[1]{$#1\times 10^{10}\,\mathrm{cm}^{-2}$\xspace}

\begin{document}

\title{Extended search for supernova-like neutrinos in NOvA coincident with LIGO/Virgo detections}

\preprint{FERMILAB-PUB-21-276-ND}

\input{authorlist.tex}

\newcommand{\plainms}{$\mathrm{M}_\odot$}
\newcommand{\ms}{\,\plainms\xspace}

\begin{abstract}

A search is performed for supernova-like neutrino
interactions coincident with 76 gravitational wave events detected by the LIGO/Virgo
Collaboration.  For 40 of these events, full readout of the time around the gravitational
wave is available from the NOvA Far Detector.  For these events, we set limits on the fluence
of the sum of all neutrino flavors of
$F < 7(4)\times 10^{10}\,\mathrm{cm}^{-2}$ at 90\% C.L. assuming energy and time distributions
corresponding to the Garching supernova models with masses 9.6(27)\,$\mathrm{M}_\odot$.
Under the hypothesis that any given gravitational wave event was caused by a supernova,
this corresponds to a distance of $r > 29(50)$\,kpc at 90\% C.L.  Weaker limits are set for
other gravitational wave events with partial Far Detector data and/or Near Detector data.

\end{abstract}

\maketitle

%\linenumbers

\section{\label{sec:introduction}Introduction}

Multimessenger astronomy is a rapidly expanding field, with exciting
opportunities to simultaneously observe violent astrophysical events using
gravitational waves (GW), electromagnetic radiation, cosmic rays, and neutrinos.
To date, a single gravitational wave
event has been associated with electromagnetic activity~\cite{ligo170817,Goldstein:2017mmi,Monitor:2017mdv},
and none have been associated with the other channels.  Not all gravitational waves and gravitational wave
candidates to date have been identified by the LIGO/Virgo Collaboration (LVC)
with a particular production mechanism~\cite{ligo2020cat}.  Although all clearly identified
events are associated with compact object mergers, there remains
the possibility that one or more was caused by a supernova,
which are expected to produce
gravitational waves, but with great uncertainty in predictions of the
signal strength~\cite{Abbott:2016tdt}.  
These potential supernovae may have evaded
optical detection either because they were obscured by dust in the central Galaxy,
or because they were ``failed'' supernovae in which the star collapsed, but did
not explode~\cite{Adams:2013ana}.

In a previous paper~\cite{mmpaper} we described a broad search for signals,
across the MeV to TeV range, associated with 26 gravitational wave events.  We
now focus on the possibility of detecting supernova-like neutrinos and present
an improved search using the now-available larger catalog of gravitational wave
events.  The paper is organized as follows.  In \sect{detectors}, we introduce
the NOvA detectors.  \sect{data} details the data set used in this analysis.
\sect{simulation} explains how we simulate supernova neutrino interactions.
\sect{analysis} describes the improved selection of supernova-like neutrinos.
Finally, \sect{results} gives the results.

\section{\label{sec:detectors}Detectors}

% doc-40273
The NOvA experiment
consists of two similar detectors, the Near Detector (ND) and the Far Detector
(FD). The ND is located at the Fermi National Accelerator Laboratory (Fermilab), 100\,m underground, while the FD is
located near Ash River, Minnesota, on the surface with a modest overburden
consisting of 1.25\,m of concrete covered with 16\,cm of barite gravel.

The NOvA detectors are segmented liquid scintillator tracking calorimeters.
Alternating planes of cells are oriented horizontally and vertically,
forming two views that can be used to reconstruct three-dimensional positions.
The cells have a cross
section of 4\,cm by 6\,cm and are 15.5\,m (3.8\,m) long in the FD (ND).  The FD
has 896 planes of cells and a total mass of 14\,kt, whereas the ND has 214
planes and a total mass of 300\,t.  The last 20 planes at the north end of the
ND are a muon catcher. They are interleaved with ten 10-cm-thick planes
of steel for the purpose of measuring the energy of muons produced in beam
interactions.  The FD has no similar structure.  The detectors are described in more detail elsewhere~\cite{novatdr}.

Light produced in the scintillator is collected by wave\-length-\allowbreak
shifting fibers and converted into electrical signals using avalanche
photodiodes.  These signals are continuously digitized at 2\,MHz at the FD and
8\,MHz at the ND.  Samples
rising above a threshold, called \emph{hits}, are retained for further
processing.  Hits from all channels are collected into 50\,$\mu$s blocks and
can be saved for offline analysis if a software trigger requests them within
about 20 minutes for the Far Detector and 30 minutes at the Near Detector.
Triggers can either be based on the content of the data or on external
signals.  Two of the latter type of triggers are used in this analysis.  First,
when LVC publishes an observation of a gravitational wave candidate over 
the Gamma-ray Coordinates Network, we respond by reading out 45\,s of
continuous data from both the ND and FD, beginning 5.16\,s prior to the
gravitational wave timestamp.  Second, we run a minimum bias pulser trigger on the FD
which reads out 550\,$\mu$s segments of data at a rate of 10\,Hz.  When only
pulser data is available, we use a window of 1000\,s centered on the
gravitational wave timestamp to match the convention established by other
neutrino observatories~\cite{Baret:2011tk,icecube1}.

The NOvA detectors are exposed to Fermilab's NuMI beam~\cite{numi}, a wideband
neutrino beam with a peak at 2\,GeV consisting mainly of either $\nu_\mu$ or
$\bar\nu_\mu$, depending on the operating mode.  Typically, the beam is
operated October through June with pulses of 10\,$\mu$s separated by 1.3\,s.
For the purposes of the analysis reported here, the beam has no impact on the
FD data since the number of beam neutrino interactions is negligible.  However,
it is a source of background at the ND; a procedure to remove beam backgrounds is
detailed in \sect{analysis}.

\section{\label{sec:data}Data Set}

\ifpdf\tableone\fi
\ifpdf\tableonetwo\fi

Tables \ref{tab:events} and \ref{tab:events2} show a summary of NOvA data collected for each of the
gravitational wave events and candidates (henceforth called ``events'') announced by LVC to date in their two
catalogs~\cite{ligocat,ligo2020cat} and via the Gamma-ray Coordinates Network~\cite{\citeuncat}.
With the exception of four gravitational wave events, at least one of the NOvA detectors
was operating and taking useful data for each event.  LVC issued public triggers beginning
with their ``O3'' run period in 2019; prior to that point, NOvA has only the FD pulser
data.  Thirteen events in O3 were only announced in the second LVC catalog
and not via public trigger; we only have FD pulser data for these as well.

% For this paper, I'm putting my foot down and using singular 'data'

Of the remaining 52 GW events that did have public triggers, we recorded
all or part of the desired 45\,s of continuous data at the FD for 32 events,
and at the ND for 40.  In five cases, the ND recorded full readouts when the 
FD did not because it has a deeper data buffer.  
At each detector, data is read out approximately in time order; alerts that arrived
when the data was near the end of the buffer resulted in partial readouts, as shown
in the table.  In the remaining three cases, the FD was down and the ND was up.

\section{\label{sec:simulation}Simulation}

Supernova neutrino interactions are simulated for use in training the selector
and for assessing signal significance.  The simulation is based on the Garching
9.6\ms and 27\ms supernova flux models~\cite{Mirizzi:2015eza}, with neutrino
interactions produced with GENIE v3.0.6~\cite{Andreopoulos:2009rq}, and the
resulting particles tracked through the detector geometry using {\sc Geant4}
v10.4.2~\cite{geant}.  The simulation only includes neutrinos above 10\,MeV,
with inverse beta decay on hydrogen (IBD) and electron elastic scattering (ES)
interactions included.  Since NOvA is hydrocarbon-based, IBD strongly dominates over
ES.  IBD is the most important interaction for NOvA because it has a large
cross section and produces a high energy positron.  The mean positron energy
produced in the 9.6(27)\ms simulation is 19.0(21.2)\,MeV.

In IBD interactions, both positrons and neutrons are simulated. Although NOvA is
primarily sensitive to positrons and electrons, the 8\,MeV of gammas from
neutron capture on $^{35}$Cl is also visible.  The NOvA detectors are 16\%
chlorine by mass.
After selection cuts, the FD has no significant sensitivity to
electrons and positrons below 10\,MeV, however, the ND is still marginally
sensitive at this energy, so the simulation somewhat undercounts
the neutrino interactions that would be selected in a real supernova.

Besides undercounting low energy neutrino interactions, the simulation also
does not include various interaction channels on carbon such as
$\nu_e\,+\,^{12}\mathrm{C} \rightarrow e^{-}\,+\,^{12}\mathrm{N}$, nor similar
channels involving other isotopes in the NOvA materials, although in many
cases, these interactions would be easily visible.  The limits set below are
therefore conservative, although IBD would dominate over these other channels
even if they were included.  We use a model without neutrino oscillations or
other flavor-changing effects because there is not enough information available
to know whether these effects would increase or decrease the number of neutrinos observed
by NOvA~\cite{Mirizzi:2015eza}.

\section{\label{sec:analysis}Analysis}

Relative to our previous report~\cite{mmpaper}, the clustering algorithm for grouping hits
into supernova neutrino event candidates has been greatly improved.  Previously, a
cluster was defined as a pair of hits with one hit in each view.
Now a cluster may have 2 to 7 hits associated in time and space.  Clusters of greater
than 7 hits are rejected as being too large to have been produced by a supernova neutrino
interaction.  In the ND, we allow clusters
with all hits in a single view.  However, at the FD, 3-dimensional position
information is essential for reducing background, so clusters must include
hits in both views.  Similarly, ND clusters may
be non-contiguous, with gaps either between hits within a detector plane or
between detector planes, but FD clusters must be contiguous to reduce
background.  Previously, we excluded the muon catcher region of the ND;
clusters in this region are now accepted.

\ifpdf\figureone\fi

Critical to the reduction of background, particularly at the FD, is the
inclusion of several new variables in the classifier that relate the distance
in time and space between candidate hit clusters and recent cosmic rays.
Michel electrons from stopping muons are a common background in the FD,
occurring at a rate of 40\,kHz.  Most Michel electrons are identified by close
association with track ends, but a small fraction of apparent Michel electrons
appear far from the track end, either because of reconstruction failures,
inefficiencies in producing hits, complex particle interactions, or some
combination of these.  Candidate clusters are judged based on their proximity
to the track end, to any point along the track, as well as to any hit in a
large cluster of activity with no reconstructed tracks.

% Do we need more detail on the cosmic rejection variables or classifier
% training here?

Supernova neutrino-like hit clusters are separated into signal and background
samples using the scikit-learn~\cite{scikit-learn} package's
\texttt{RandomForestClassifier} class.  The classifier is trained with
simulated 9.6\ms supernova interactions and real minimum-bias data from the NOvA
detectors.  The classifier was optimized separately for the ND and the FD.
Further, it is optimized separately for the two cases of FD data ---
continuous readout and pulser.  The pulser data must be treated
differently because the look-back time for cosmic rays that may have produced a
background cluster is reduced.  Additionally, since the livetime is smaller,
efficiency is prioritized over background reduction.  In
all three cases, the figure of merit~\cite{Punzi:2003bu}
\[\frac{\mathrm{signal}}{a/2 + \sqrt{\mathrm{background}}},\] is optimized,
with $a=1.292$ to optimize 90\% C.L. limits.  The resulting efficiencies for
IBD positrons are shown in \fig{positroneff}.  Efficiencies for ES electrons, as
a function of electron energy, are very similar.  At the ND, neutron captures
from IBD are selected with 2\% efficiency, while at the FD the neutron capture
efficiency is negligible for purposes of the signal: only 0.02\%.  No attempt
is made in the analysis to associate positron and neutron delayed coincidences
in either detector.

Compared with our previous analysis method, the rate of selected background
candidates in the FD, for continuous readout, is reduced by a factor of eighty,
from 460\,Hz to 6\,Hz, while the signal efficiency for IBD positrons is reduced from 7.8\% to
4.3\%.  The reduced signal efficiency is a consequence of the optimization described above.  In the previous analysis, the same selection was used for FD continuous-readout
and FD pulser data.  In this analysis, the pulser background rate is reduced to 
55\,Hz, while the signal efficiency is increased to 9.0\%, or 0.3\,Hz and 0.05\%
taking into account the 0.55\% livetime.
In the ND, the rate of selected background candidates has been slightly reduced
from 0.5\,Hz to 0.4\,Hz while the signal efficiency has been increased from
12\% to 44\%.

Since the neutrino event classifier is trained on real detector data, no
explicit identification of the background components is made.  The FD
background likely contains significant components from cosmogenic thermal
neutron captures, cosmogenic $^{12}$B and $^{12}$N beta decays, and single-hit
uranium/thorium-chain radioactivity paired with unrelated single-hit
electronics noise.  The latter is possibly a significant component of the ND
background as well, but cosmogenic activity is strongly suppressed compared to
the FD.

For 16 of the 40 GW events with ND data, the NuMI beam was in operation.  Data
at the ND is taken in 5\,ms segments.  Any 5\,ms data segment is rejected if it
overlaps with a beam pulse or the time up to 3\,ms following a beam pulse.
This conservative cut removes all prompt beam activity, muon decays, and
neutron captures from thermal neutrons that were produced in the detector and
remained in the detector until captured.  Some neutron captures can be delayed
up to several milliseconds if thermal neutrons spend time in the air
surrounding the detector; the 3\,ms cut rejects a large majority of these neutrons.  

For each gravitational wave event, we first examine the selected clusters in
1-second bins searching for any significant excess over background, where the
background level is determined {\it in situ} from the 45\,s readout (or 1000\,s
window in the case of FD pulser data).  Second, we assume that a supernova
burst begins at the gravitational wave timestamp and set limits on its strength
for the case of the Garching 9.6\ms and 27\ms models.  Because NOvA's
efficiency rises rapidly with neutrino energy between 10 and 30\,MeV, the
higher neutrino energies in the 27\ms model result in stronger fluence limits.

Depending on the state of the two NOvA detectors and whether a trigger was
received from LVC, several different types of data sets can be available.  The
best case is when a timely trigger was received and we read 45\,s of continuous
data from the ND and FD.  In this case, a joint analysis is done using the data
from the two detectors.  The FD provides more statistical power, but the ND
still makes a significant contribution.  In some cases, continuous data is
available from the ND, but only pulser data from the FD.  Again, a joint
analysis is performed, but in this case, the ND provides nearly all the
statistical power.  In some cases, the continuous data from ND or FD is not a
complete 45\,s, but in all those cases enough was read out to establish the
background level and allow the analysis to be run without modification.  The
background level is not determined with as much precision in these cases,
leading to a slight weakening of limits.  Finally, in some cases, data from
only one detector is available.  The status for all GW events is shown in
\tab{events}.

\section{\label{sec:results}Results}

\ifpdf \figuretwo \fi

No excess over background is observed for any gravitational wave event at any
time within the analyzed window.  Background rates were stable at both
detectors, being around 5\,Hz at the FD for the continuous-readout selection,
0.3\,Hz for the FD pulser selection and 0.4\,Hz at the ND.  Assuming all
selected clusters are background, the limits depend on statistical fluctuations
in the background in the first few seconds after the gravitational wave
timestamp.  A typical event is shown in \fig{timedists}.  

For each GW event, 90\% C.L. limits are set on the fluence of the sum of all
neutrino flavors, $\flusym$, under the assumption of the two Garching supernova models discussed
above, without flavor-changing effects.  The limits are set via a fit to the
time series of neutrino candidates with two parameters: the background rate and
the signal strength, with the signal templates as shown in \fig{timedists}.  A
Bayesian approach is used with flat priors in each parameter.  

A posterior PDF, profiled over the background level, is constructed by scanning
over signal strength, relative to the prediction at 10\,kpc, in steps of
$10^{-3}$.  At each step, the binned log-likelihood, \[-\log L =
\sum_\mathrm{i}\left( m_i - d_i + d_i\log\frac{d_i}{m_i}\right)\] is computed for the
background normalization that minimizes $-\log L$, where $m_i$ is the number of
events predicted by the model in bin $i$ and $d_i$ is the number of events observed, likewise.  The
probability density is proportional to $L$.  The resulting curve is integrated
numerically up to 90\% of the total, and this sets the 90\% upper limit on
signal strength, $s_{90}$.   Because the signal would decrease as $1/r^2$, where $r$
is the distance to the hypothetical supernova, the 90\% lower limit on distance is $r_{90} =
10\,\mathrm{kpc}/\sqrt{s_{90}}$.  Given the number of neutrinos predicted by
each model, $N = 6.8 (11) \times 10^{57}$ for 9.6(27)\ms, fluence limits,
$F_{90}$, are related to the distance limits via \[F_{90} = \frac{N}{4\pi
r_{90}^2}.\]

No systematic
effects are explicitly included in the procedure, but as detailed in
\sect{simulation}, our estimate of the rate of detectable neutrino interactions
is conservative; we believe this conservatism is sufficient to cover any
systematic effects in signal efficiency.  All limits are shown in \tab{events}
and a discussion of notable features thereof follows. 

For gravitational wave events in which we read out continuous FD data in
response to an LVC trigger (the best case), fluence limits range between
\flu{4} and \flu{15}, assuming the 9.6\ms Garching model.  In this model, 22\%
of the neutrinos are $\bar\nu_\mathrm{e}$, to which NOvA is primarily
sensitive.  The median limit is \flubare{7}.  Similarly, for the 27\ms model,
in which 23\% of the flux is $\bar\nu_\mathrm{e}$, we set limits ranging from
\flu{2.1} to \flu{9} with a median of \flubare{4}.  If interpreted as limits on
the distance to a hypothetical supernova, we exclude a 9.6\ms supernova in the
median case, at 90\% C.L., closer than 29\,kpc.  For the event with the
strongest exclusion, S200115j, we exclude a 9.6\ms supernova closer than
40\,kpc.  For the 27\ms model, we exclude a supernova, in the median case,
closer than 50\,kpc, and for S200115j, 70\,kpc.

In the next best case, we have continuous ND data, but either have no FD data
or only pulser data from the FD data.  In the latter case, the limit is
strongly dominated by the ND data.  Fluence limits for the 9.6\ms model range
from \flu{19} to \flu{29}, and for the 27\ms from \flu{13} to \flu{22}.
Because of the ND's lower background, the efficiency for selecting lower energy neutrinos is
higher than the FD.
The flux model therefore has less effect on fluence limits dominated by ND data.  The median
distance limit for a 9.6(27)\ms supernova is 16(24)\,kpc.

Finally, when using only FD pulser data, fluence limits range from \flu{190} to
\flu{700} for the 9.6\ms model and from \flu{110} to \flu{400} for the 27\ms
model.  Even with only FD pulser data, some exclusion of supernovae in or
behind the Galactic core (at $\sim 8\,$\,kpc), whose optical signal may have
been obscured, is possible, with distance limits ranging from 2.9--6\,kpc for
the 9.6\ms case and 5--9\,kpc for the 27\ms case.

The 26 GW events analyzed in our previous report are reanalyzed using the
improved analysis.  The limits quoted for the seven previously-analyzed
events that include FD and/or ND continuous readout are now stronger, in the
median case, by a factor of three, and in no case is the result we now give
weaker than our previously published result.  However, for events with only FD
pulser data, the new analysis techniques only yield a 40\% improvement in
fluence limits.  There are four GW events that, in the new analysis, have a
weaker limit for at least one of the two supernova models, GW170608, GW170729,
GW170823 and GW190521\_074359.  This is an expected consequence of using an
analysis that is almost entirely different than our previous analysis, such
that there is little correlation between the hits selected previously and now.

\section{\label{sec:conclusions}Conclusions}

We have searched for supernova-like neutrinos coincident with
76 gravitational wave events reported by LVC.  No excess consistent with
such neutrinos is found.  Assuming a burst of supernova-like neutrinos
beginning at LVC's reconstructed gravitational wave time, we set limits
on the fluence of supernova-like neutrinos under two supernova models.
In the 32 cases with full FD data, these limits are sufficient to
largely exclude the possibility that any of the gravitational waves
originated from a stellar core collapse in our galaxy.  This includes
the ``failed supernovae'' in which there is no explosion and/or
scenarios that lead to early black hole formation, since similar neutrino
luminosities are expected in any of these cases~\cite{Schneider2020,Li:2020ujl}. 
Our search complements those performed by other neutrino observatories~\cite{borexino,kamland,kamland2,superk,superk2,icecube1,icecube2,pa}.  
The NOvA detectors will
continue to operate for several years, including during the
upcoming O4 run of LVC.

This document was prepared by the
NOvA collaboration using the resources of the Fermi National Accelerator
Laboratory (Fermilab), a U.S. Department of Energy, Office of Science, HEP User
Facility. Fermilab is managed by Fermi Research Alliance, LLC (FRA), acting
under Contract No. DE-AC02-07CH11359. This work was supported by the U.S.
Department of Energy; the U.S. National Science Foundation; the Department of
Science and Technology, India; the European Research Council; the MSMT CR, GA
UK, Czech Republic; the RAS, RFBR, RMES, RSF, and BASIS Foundation, Russia;
CNPq and FAPEG, Brazil; STFC, UKRI, and the Royal Society, United Kingdom; and the
State and University of Minnesota.
We are grateful for the contributions of the staffs of the
University of Minnesota at the Ash River Laboratory and of Fermilab.

% DON'T give bibliographystyle here; it's already been set by the documentclass options
\ifpdf
  \bibliography{smmpaper-arxiv}
\else
  \bibliography{smmpaper}
  \clearpage
  \tableone
  \tableonetwo
  \figureone
  \figuretwo
\fi

%%%%%%%%%%%%%%%%%%%%%%%%%%%%%%%%%%%%%%%%%%%%%%%%%%%%%%%%%%%%%%%%%%%%%%%%
\end{document}

%% file: authorlist.tex
%NOvA 2nd Multimessenger paper
%last Version 8 June 2021, with legacy authors Maintained by Maury Goodman
\newcommand{\ANL}{Argonne National Laboratory, Argonne, Illinois 60439, 
USA}
\newcommand{\ICS}{Institute of Computer Science, The Czech 
Academy of Sciences, 
182 07 Prague, Czech Republic}
\newcommand{\IOP}{Institute of Physics, The Czech 
Academy of Sciences, 
182 21 Prague, Czech Republic}
\newcommand{\Atlantico}{Universidad del Atlantico,
Carrera 30 No. 8-49, Puerto Colombia, Atlantico, Colombia}
\newcommand{\BHU}{Department of Physics, Institute of Science, Banaras 
Hindu University, Varanasi, 221 005, India}
\newcommand{\UCLA}{Physics and Astronomy Department, UCLA, Box 951547, Los 
Angeles, California 90095-1547, USA}
\newcommand{\Caltech}{California Institute of 
Technology, Pasadena, California 91125, USA}
\newcommand{\Cochin}{Department of Physics, Cochin University
of Science and Technology, Kochi 682 022, India}
\newcommand{\Charles}
{Charles University, Faculty of Mathematics and Physics,
 Institute of Particle and Nuclear Physics, Prague, Czech Republic}
\newcommand{\Cincinnati}{Department of Physics, University of Cincinnati, 
Cincinnati, Ohio 45221, USA}
\newcommand{\CSU}{Department of Physics, Colorado 
State University, Fort Collins, CO 80523-1875, USA}
\newcommand{\CTU}{Czech Technical University in Prague,
Brehova 7, 115 19 Prague 1, Czech Republic}
\newcommand{\Dallas}{Physics Department, University of Texas at Dallas,
800 W. Campbell Rd. Richardson, Texas 75083-0688, USA}
\newcommand{\DallasU}{University of Dallas, 1845 E 
Northgate Drive, Irving, Texas 75062 USA}
\newcommand{\Delhi}{Department of Physics and Astrophysics, University of 
Delhi, Delhi 110007, India}
\newcommand{\JINR}{Joint Institute for Nuclear Research,  
Dubna, Moscow region 141980, Russia}
\newcommand{\Erciyes}{
Department of Physics, Erciyes University, Kayseri 38030, Turkey}
\newcommand{\FNAL}{Fermi National Accelerator Laboratory, Batavia, 
Illinois 60510, USA}
\newcommand{\UFG}{Instituto de F\'{i}sica, Universidade Federal de 
Goi\'{a}s, Goi\^{a}nia, Goi\'{a}s, 74690-900, Brazil}
\newcommand{\Guwahati}{Department of Physics, IIT Guwahati, Guwahati, 781 
039, India}
\newcommand{\Harvard}{Department of Physics, Harvard University, 
Cambridge, Massachusetts 02138, USA}
\newcommand{\Houston}{Department of Physics, 
University of Houston, Houston, Texas 77204, USA}
\newcommand{\IHyderabad}{Department of Physics, IIT Hyderabad, Hyderabad, 
502 205, India}
\newcommand{\Hyderabad}{School of Physics, University of Hyderabad, 
Hyderabad, 500 046, India}
\newcommand{\IIT}{Illinois Institute of Technology,
Chicago IL 60616, USA}
\newcommand{\Indiana}{Indiana University, Bloomington, Indiana 47405, 
USA}
\newcommand{\INR}{Institute for Nuclear Research of Russia, Academy of 
Sciences 7a, 60th October Anniversary prospect, Moscow 117312, Russia}
\newcommand{\Iowa}{Department of Physics and Astronomy, Iowa State 
University, Ames, Iowa 50011, USA}
\newcommand{\Irvine}{Department of Physics and Astronomy, 
University of California at Irvine, Irvine, California 92697, USA}
\newcommand{\Jammu}{Department of Physics and Electronics, University of 
Jammu, Jammu Tawi, 180 006, Jammu and Kashmir, India}
\newcommand{\Lebedev}{Nuclear Physics and Astrophysics Division, Lebedev 
Physical 
Institute, Leninsky Prospect 53, 119991 Moscow, Russia}
\newcommand{\Magdalena}{Universidad del Magdalena, Carrera 32 No 22-08 Santa Marta, Colombia}
\newcommand{\MSU}{Department of Physics and Astronomy, Michigan State 
University, East Lansing, Michigan 48824, USA}
\newcommand{\Crookston}{Math, Science and Technology Department, University 
of Minnesota Crookston, Crookston, Minnesota 56716, USA}
\newcommand{\Duluth}{Department of Physics and Astronomy, 
University of Minnesota Duluth, Duluth, Minnesota 55812, USA}
\newcommand{\Minnesota}{School of Physics and Astronomy, University of 
Minnesota Twin Cities, Minneapolis, Minnesota 55455, USA}
\newcommand{\Mississippi}{University of Mississippi, University, Mississippi 38677, USA}
\newcommand{\NISER}{National Institute of Science Education and Research,
Khurda, 752050, Odisha, India}
\newcommand{\Oxford}{Subdepartment of Particle Physics, 
University of Oxford, Oxford OX1 3RH, United Kingdom}
\newcommand{\Panjab}{Department of Physics, Panjab University, 
Chandigarh, 160 014, India}
\newcommand{\Pitt}{Department of Physics, 
University of Pittsburgh, Pittsburgh, Pennsylvania 15260, USA}
\newcommand{\QMU}{School of Physics and Astronomy,
 Queen Mary University of London,
London E1 4NS, United Kingdom}
\newcommand{\RAL}{Rutherford Appleton Laboratory, Science 
and 
Technology Facilities Council, Didcot, OX11 0QX, United Kingdom}
\newcommand{\SAlabama}{Department of Physics, University of 
South Alabama, Mobile, Alabama 36688, USA} 
\newcommand{\Carolina}{Department of Physics and Astronomy, University of 
South Carolina, Columbia, South Carolina 29208, USA}
\newcommand{\SDakota}{South Dakota School of Mines and Technology, Rapid 
City, South Dakota 57701, USA}
\newcommand{\SMU}{Department of Physics, Southern Methodist University, 
Dallas, Texas 75275, USA}
\newcommand{\Stanford}{Department of Physics, Stanford University, 
Stanford, California 94305, USA}
\newcommand{\Sussex}{Department of Physics and Astronomy, University of 
Sussex, Falmer, Brighton BN1 9QH, United Kingdom}
\newcommand{\Syracuse}{Department of Physics, Syracuse University,
Syracuse NY 13210, USA}
\newcommand{\Tennessee}{Department of Physics and Astronomy, 
University of Tennessee, Knoxville, Tennessee 37996, USA}
\newcommand{\Texas}{Department of Physics, University of Texas at Austin, 
Austin, Texas 78712, USA}
\newcommand{\Tufts}{Department of Physics and Astronomy, Tufts University, Medford, 
Massachusetts 02155, USA}
\newcommand{\UCL}{Physics and Astronomy Department, University College 
London, 
Gower Street, London WC1E 6BT, United Kingdom}
\newcommand{\Virginia}{Department of Physics, University of Virginia, 
Charlottesville, Virginia 22904, USA}
\newcommand{\WSU}{Department of Mathematics, Statistics, and Physics,
 Wichita State University, 
Wichita, Kansas 67206, USA}
\newcommand{\WandM}{Department of Physics, William \& Mary, 
Williamsburg, Virginia 23187, USA}
\newcommand{\Wisconsin}{Department of Physics, University of 
Wisconsin-Madison, Madison, Wisconsin 53706, USA}
\newcommand{\deceased}{Deceased.}
\affiliation{\ANL}
\affiliation{\Atlantico}
%\affiliation{\Athens}
\affiliation{\BHU}
%\affiliation{\UCLA}
\affiliation{\Caltech}
\affiliation{\Charles}
\affiliation{\Cincinnati}
\affiliation{\Cochin}
\affiliation{\CSU}
\affiliation{\CTU}
%\affiliation{\DallasU}
\affiliation{\Delhi}
\affiliation{\Erciyes}
\affiliation{\FNAL}
\affiliation{\UFG}
\affiliation{\Guwahati}
\affiliation{\Harvard}
\affiliation{\Houston}
\affiliation{\Hyderabad}
\affiliation{\IHyderabad}
\affiliation{\IIT}
\affiliation{\Indiana}
\affiliation{\ICS}
\affiliation{\INR}
\affiliation{\IOP}
\affiliation{\Iowa}
\affiliation{\Irvine}
%\affiliation{\Jammu}
\affiliation{\JINR}
\affiliation{\Lebedev}
\affiliation{\Magdalena}
\affiliation{\MSU}
%\affiliation{\Crookston}
\affiliation{\Duluth}
\affiliation{\Minnesota}
\affiliation{\Mississippi}
%\affiliation{\Oxford}
\affiliation{\NISER}
\affiliation{\Panjab}
\affiliation{\Pitt}
\affiliation{\QMU}
%\affiliation{\RAL}
\affiliation{\SAlabama}
\affiliation{\Carolina}
%\affiliation{\SDakota}
\affiliation{\SMU}
\affiliation{\Stanford}
\affiliation{\Sussex}
\affiliation{\Syracuse}
\affiliation{\Tennessee}
\affiliation{\Texas}
%\affiliation{\Dallas}
\affiliation{\Tufts}
\affiliation{\UCL}
\affiliation{\Virginia}
\affiliation{\WSU}
\affiliation{\WandM}
%\affiliation{\Winona}
\affiliation{\Wisconsin}

%Added June 2017
\author{M.~A.~Acero}
\affiliation{\Atlantico}

\author{P.~Adamson}
\affiliation{\FNAL}

%For first publication
%\author{C.~Ader}
%\affiliation{\FNAL}

%added August 2019, Legacy October 2019
%\author{G.~Agam}
%\affiliation{\IIT}

%Added March 2017
\author{L.~Aliaga}
\affiliation{\FNAL}

%Legacy July 2020
%\author{T.~Alion}
%\affiliation{\Sussex}

%Legacy July 2020
%\author{V.~Allakhverdian}
%\affiliation{\JINR}

%Legacy August 2019
%\author{S.~Altakarli}
%\affiliation{\WSU}

%Legacy June 2017
%\author{D.~Ambrose}
%\affiliation{\Minnesota}

%For first publication
%\author{M.~Andrews}
%\affiliation{\FNAL}

\author{N.~Anfimov}
\affiliation{\JINR}

%Went Legacy 1 Feb 2016
%\author{I.~Anghel}
%\affiliation{\Iowa}
%\affiliation{\ANL}

\author{A.~Antoshkin}
\affiliation{\JINR}

%For first publication
%\author{K.~Arms}
%\affiliation{\Minnesota}

% Legacy June 2017; rejoined February 2019
\author{E.~Arrieta-Diaz}
\affiliation{\Magdalena}
%\affiliation{\SMU}

%added August 2019
\author{L.~Asquith}
\affiliation{\Sussex}

%Removed June 2017
%\author{K.~Augsten}
%\affiliation{\CTU}

\author{A.~Aurisano}
\affiliation{\Cincinnati}

%For first publication
%\author{D.~S.~Ayres}
%\affiliation{\ANL}

\author{A.~Back}
\affiliation{\Iowa}

\author{C.~Backhouse}
%\affiliation{\Caltech}
\affiliation{\UCL}

% ...

\collaboration{The NOvA Collaboration}
\noaffiliation